# Super-capacitors interfaced with quantum dots at the electrolyte/electrode interface: capacitance gain and fluorescence line-width narrowing

H. Grebel

Center for Energy Efficiency, Resilience and Innovation (CEERI) and the ECE Department at the New Jersey Institute of Technology, Newark, NJ 07102. grebel@njit.edu.
https://orcid.org/0000-0001-9367-8117

**Abstract:**

Quantum dots (QDs), embedded in supercapacitors (S-C) cells exhibited capacitance amplification that peaked at certain QD concentration. More than 2.5 peak amplification was demonstrated with cyclic voltammetry (C-V) at a scan rate of 0.1 V/s. The mass ratio of the dots to that of the active-carbon electrode (A-C) was less than 1:5000. Fluorescence signals from dry (without electrolyte) and wet (with electrolyte) samples, exhibited a correlated, substantial peak enhancement vs QD concentration, as well. In the case of wet samples, fluorescence line narrowing was demonstrated, which indicates a substantial fluorescence gain. Both effects are attributed to local field effects through formation of a colloidal array at the electrode/electrolyte interface. Embedding QDs at the electrolyte/electrode interface has an added value that it may be further enhanced by white light and indeed this is the case here. Probing the electrolyte/electrode interface with fluorescing materials adds to our basic knowledge of the interface and could be useful for light-sensitive S-C cells.

## I. Introduction

Super-capacitors (S-C) are used in a wide-range of applications, such as consumer electronic products, memory back-up devices, hybrid electric vehicles, and power supplies [1-2]. They were also proposed as buffers to highly fluctuating power grids with sustainable sources [3]. S-C take an advantage of the large capacitance at the narrow interface between a porous electrode and an electrolyte [4-7]. Thus, large effect on the cell's polarity may be achieve by adding electrical dipoles to the interface. In the case of semi conductive colloids, randomly dispersed at the electrolyte/electrode interface, additional photo and thermal effects may be observed [8-9] albeit at a relatively large concentration of the fluorophores.

In the experiments, we added a low dispersion of functionalized semiconductor quantum dots (QDs) – the mass ratio between the QDs and the active carbon electrode mass was less than 1 to 5000. This resulted in peak capacitance enhancement of ~2.5 when using cyclic voltammetry at a scan rate of 0.1 V/s. The QDs were functionalized by a ligand that prevented agglomeration while in suspension. This ensured a strong electrostatic bond between them and the conductive active-carbon (A-C) electrode.

Generally, incorporating conductive features into a dielectric material that separates two electrode increases the polarization of ordinary capacitors [10-14]. This was true at the low frequency regime of super-capacitors [15] and also true at the high-frequency regime [16]. Metal colloids exhibit Raman enhancement (Surface Enhanced Raman Spectroscopy, SERS [17-21]) and IR signals enhancement (Surface Enhanced IR Absorption, SEIRA, [22-23]). Periodic structures for SERS applications have been also investigated [24]. Therefore, it makes sense to study semiconductor dots and attempt to arrange them in periodic structures.

Dispersions of the QDs are well known. As mentioned above, the QDs are coated with a ligand to prevent agglomeration while in suspension. In general, the induced charge in the conductive, or semiconductor colloids is larger for larger colloids [25]. On the other hand, the effect of placing them at the electrolyte/electrode interface is limited by the colloid conductivity and the dimensions of the related double layer (which is of order of 1 nm) and to some extent, the dimension of the diffusion layer. Using a full-wave analysis, the closer the colloids are to the conductive electrode, the larger is the super-capacitive effect [26].

Imbedding fluorescing materials in S-C have been investigated for basic science [27], energy harvesting [28] and sensing [29]. What is missing, is the role of interaction among QDs and between the QDs and the nearby electrode. The purpose of the paper is to show that the presence of well separated QDs, arranged in an array (namely with a specific lattice constant), enhances the overall S-C capacitance and also affects the fluorescence line of the dots themselves. It should be noted that in periodic structures that are at resonance with the fluorescence line, amplify the fluorescence signal and exhibit line narrowing [30].

## II. Materials

### II.a Preparation methods - QDs:

Core/shell QDs, CdSe/ZnS, were purchased from Mesolight (China) and were diluted to a concentration of 1 mg/mL in toluene. Titration of the various samples was made from this stock solution with the help of a micro-pipette directly into the prepared slurry. Vials, each containing 2 mL of the slurry (active carbon, A-C and QDs) suspended in toluene were prepared.

**II.b. Preparation methods - the porous electrodes:**

A 100 mg of poly(methyl methacrylate) was first dissolved in 20 mL of toluene. 2 g of active-carbon (A-C, specific surface area of 1100 $m^2/g$, with an average dimension of 15 $\mu m$ and produced by General Carbon Company, GCC, Paterson, NJ, USA, [8-9]) were added and sonicated for 1 hour using a horn antenna. Vials, each containing 2 mL of the slurry were prepared. To these, succession amounts of 10 $\mu L$ of QDs, suspended in toluene were added. Each mixture was further sonicated with the horn antenna for additional 30 min. The slurry was drop-casted on grafoil electrodes (area of contact 1.27x1.27 $cm^2$, manufactured by Miseal and purchased through Amazon), baked on a hot plate at <90 $^oC$ and then soaked with an electrolyte (1 M of $Na_2SO_4$). A fiberglass filter (Whatmen 1851-055), or a paper filter were used as a membranes.

**II.c. The samples:**

Cuts of 200 micron thick grafoil electrodes with back adhesive (Width=1.27 cm x Length=2.54 cm) were placed on similar cut microscope slides. Before placing it on the slides, the grafoil electrodes were heated for a few hours. Two slides were held by tweezers (or plastic clips) and the boundaries of the sample were left unsealed while soaking it in the electrolyte. The samples were later sealed with an epoxy. A hole was cut through the upper electrode and membrane to let the laser beam through. The sample configuration is shown in Fig. 1a and its picture in Fig.1b. The cell covered only half of the sample area, namely, 1.27 cm x 1.27 cm.

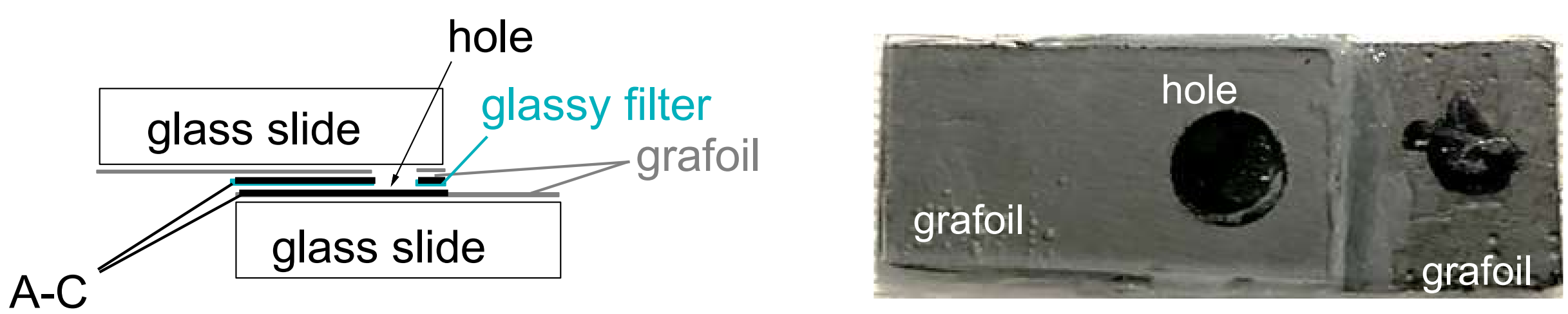

(a) (b)

Figure 1. (a) A cross section of sample with grafoil coated active-carbon (A-C) electrodes, and (b) a picture of the sealed sample, covering area of ca, 1.27x1.27 $cm^2$ with an entrance hole.

**II.d. Electrochemical Techniques:**

Potentiostat/Galvanostat (Metrohm) was used in a 2-electrode set-up. Data is provided with Cyclic Voltammetry (C-V) at a scan rates of 0.1 V/s. Data was accumulated for 10 scans of which only the last one is shown. The data was collected without light and with a 75 W white light that was placed at 30 cm from the sample surface. Obviously, not all the source spectrum is associated with the illumination effect [8-9].

**II.e. Fluorescence Measurements:**

The fluorescence (FL) system was composed of a 22-cm spectrometer equipped with a camera (cooled to -15 $^oC$). A Coherent laser, at $\lambda$=488 nm, with intensity of I=5 mW at the sample and x10 objective were used. Measurements were made with titrated concentrations of the QDs on dry substrates and when the S-C were filled with electrolyte (wet sample). For the latter, a hole was made in the upper electrode to let the excitation laser through. Data was collected for 30 s with the aid of a computer and was fitted with Gaussian curves. Fluorescence data was also collected with a 532 nm laser at I=5 mW on the sample surface, as well. Since the fluorescence line is not as sensitive to the excitation laser line one may conclude that the observed effects are related to the excited QD line (at 585 nm). Also note that the PMMA binder is transparent to the excitation and the fluorescence lines.

## III. Results

### III.a. Cells capacitance

Cyclic voltammetry (CV) curves for a reference (no QDs) and one of the samples that does contain QDs are shown in Fig. 2a-b. The scan rate was 0.1 V/s. The black curves are for the cases without white-light illumination, whereas the red curves were obtained with a 75 W incandescent white light situated at 30 cm from the surface.

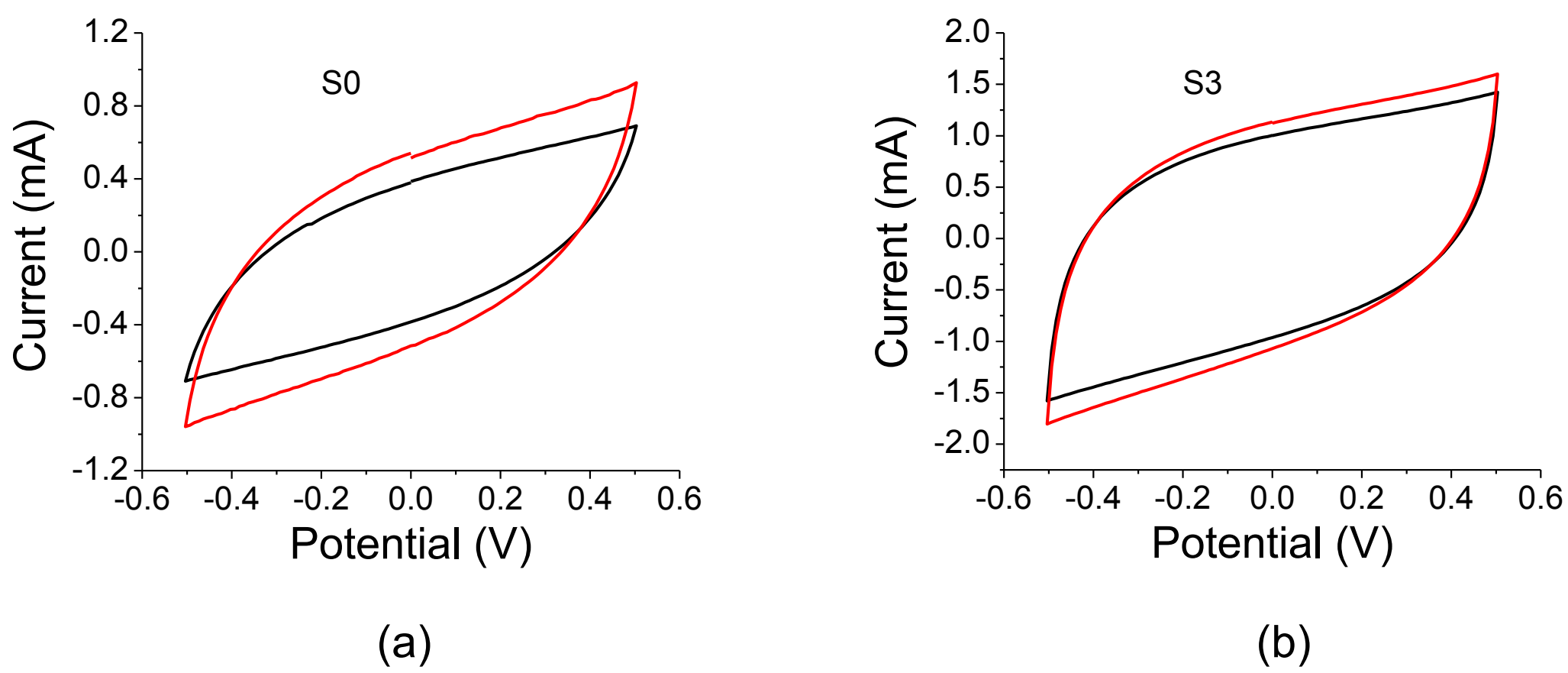

Figure 2. C-V curves for a reference sample (a, no QDs) and (b, the sample with increased capacitance). The scan rate was 0.1 V/s. The black curve were obtained without light while the red curves were obtained with a 75 W white light place at 30 cm from the sample surface.

The gravitonic specific capacitance (in F/g) as a function of concentration is shown in Fig. 3a. A clear capacitance peak appears between 30 to 40 μg of QDs when incorporated in 200 mg of A-C mass of the electrode (or, explicitly, 30 or 40 μg of QDs in samples containing 200 mg of A-C). We reiterate that the 200 mg of A-C and the QDs were immersed in vials containing 2 mL of toluene and the original batch concentration of the QDs was 1 mg/mL. As argued later, the peak signifies the onset of optimal local field due to the formed array of QDs and the points to a transition from a quasi-periodic QDs array to an amorphous phase where all QDs are randomly pile up on top of each other (see the Discussion Section).

Plot of the absolute difference between the illuminated and non-illuminated cased is shown in Fig. 3b. Almost constant effect is noted while the QD concentration is increased up to 40 μg. A substantial decrease occurs beyond that value.

The area of the I-V curve (used in the calculation of the gravitonic capacitance) is shown in Fig. 3c. Since the area of all samples is similar, this curve provides to areal capacitance, ascertaining quasi 2-D effects. Here as well, one may observe a capacitance peak at 40 μg. As a note, the A-C electrode is of a fractal nature with a dimension of ca 2.5.

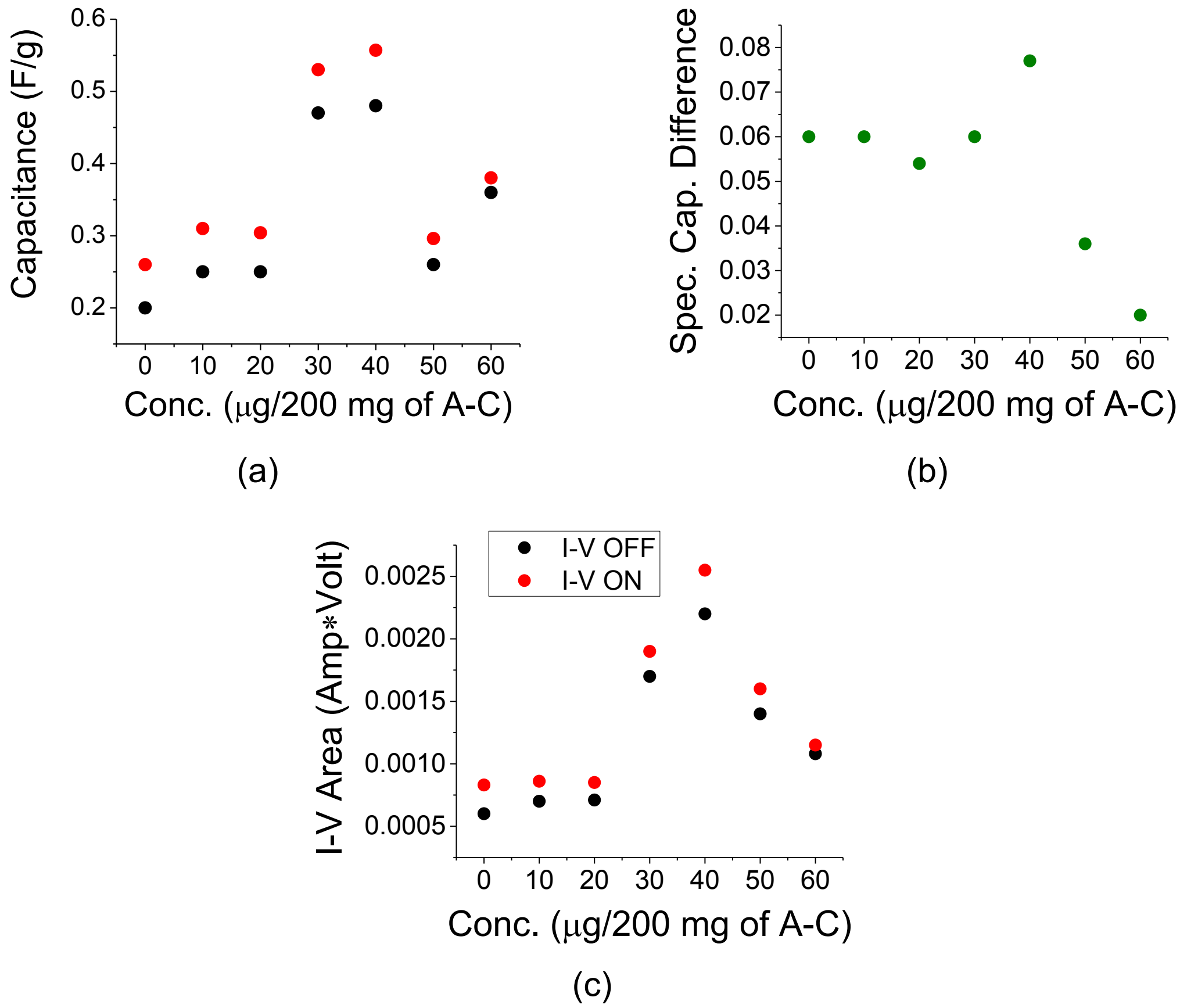

Figure 3. (a) Gravitonic specific capacitance in F/g as a function QDs concentration quoted with respect to the mass of the mass of the active carbon (A-C). Black dots: no illumination; Red dots: with white-light illumination. (b) Absolute change of the specific capacitance ($C_{ON}$-$C_{OFF}$) as a function of dot concentration. (c) Plot of the I-V curve area as a function of QD concentration, with (I-V ON) and without white-light illumination (I-V OFF).

Figs. 4 shows the relative effect of light illumination on the capacitance value, ($C_{ON}$-$C_{OFF}$)/$C_{OFF}$, where ON and OFF refer to the measurement with and without light, respectively. The curve portray monotonous and declining trend with an exception of a small dip at QDs concentration of 30 μg/200 mg of A-C. The small dip may signify the onset of local field by the resonating array of QDs [16]; since the illumination effect is constant, the relative capacitance is decreased due to the large capacitance increase at this point. At higher QDs concentrations, the electrode was masked by the semiconductor QDs layer, and the relative optical effect has diminished.

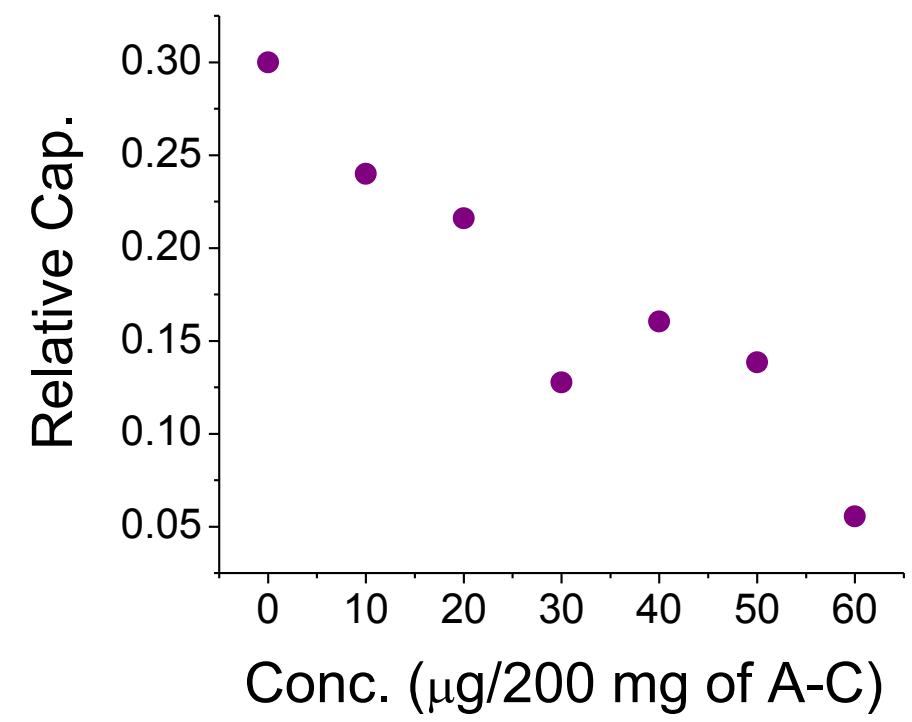

Figure 4. Trend of the relative gravitonic specific capacitance, $(C_{ON}-C_{OFF})/C_{OFF}$, as a function of QDs concentration.

### III.b. Fluorescence Measurements

Dry samples: fluorescence data are presented in Fig. 5a for dry samples. CdSe/ZnS QDs on dry substrates exhibited almost constant fluorescence line-width, ~27 nm. The fluorescence amplitude exhibited a resonance like features that peaked at 50 μg/200 mg of A-C (see comments in the discussion section). This resonating effect could be attributed to the average distance between the QDs. We point that data was randomly taken within the 1.27 cm x 1.27 cm samples using x10 objective. The use of a relatively low-power objective was aimed at averaging a relatively large collection area while minimizing local concentration effects. Local dispersion fluctuations might affect the data when a more powerful objective lens is used. One may observe two major lines: one at 585 nm - the fluorescing QDs line - and one at 660 nm. The latter is due to the A-C and grafoil substrate and appears even without the presence of the QDs. The 660 nm line may be attributed to oxide contamination of the A-C/grafoil substrates.

Wet samples: The wet samples portrayed signal enhancement as their dry sample counterpart. In addition, they also exhibited a well-defined fluorescence line narrowing at 30 μg/200 mg of A-C (Fig. 5b). The line-width at this point was smaller than to the line-width obtained for the dry samples - 22 nm vs 27 nm, respectively. Light narrowing is indicative of fluorescence gain and may point to a larger coupling between dots in the wet environment.

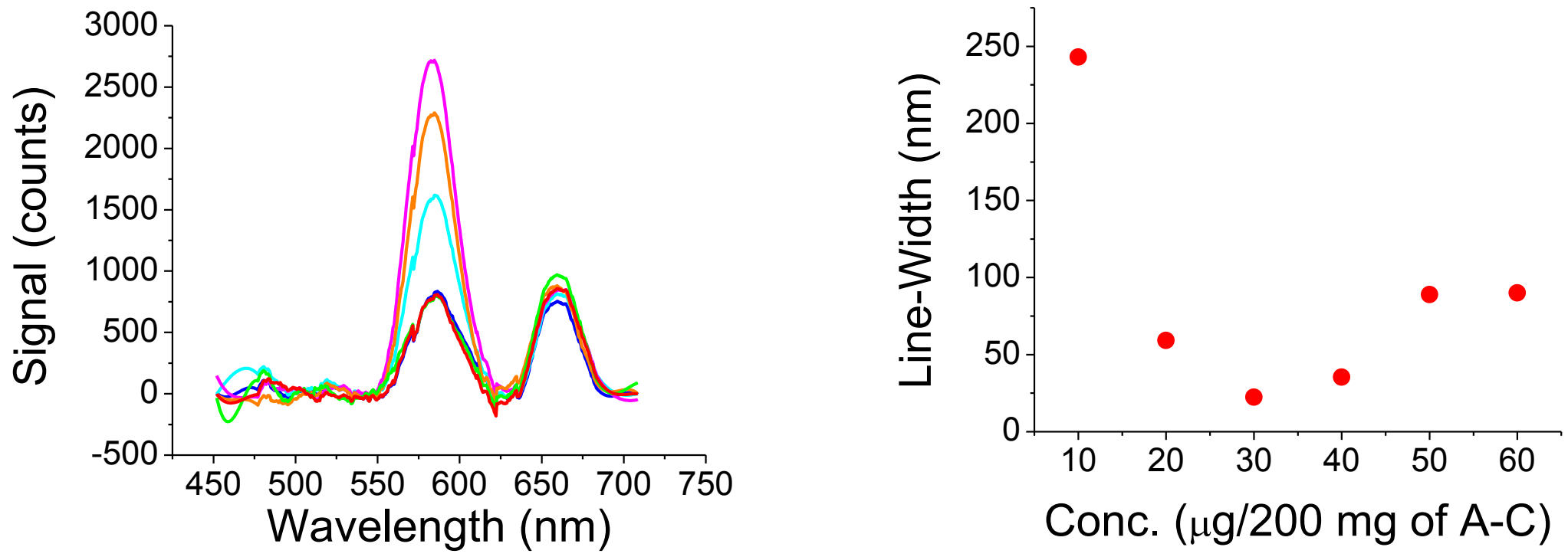

(a) (b)

Figure 5. (a) Fluorescence curves for dry samples: no substantial width-change in the fluorescence curves, despite the large amplitude increase at 50 μg/200 mg of A-C. The curves are labeled red, green, blue, cyan, purple and orange for the respective concentration: 10, 20, 30, 40, 50, 60 μg/200 mg of A-C. (b) Wet samples: line-width change as a function of QDs concentration with clear narrowing at 30 μg/200 mg of A-C.

## IV. Discussion

Capacitance changes as a function of very small concentration of QDs have been observed. We attribute the large effect to the presence of an array of QDs at close proximity to the porous electrode. As pointed earlier, most of the S-C cell's polarity comes from the electrolyte/electrode interface while the remaining of the cell is neutral. Thus, it is natural to assume that the capacitive effect is due to dots within the double- and the diffusion layers next to the electrode [26]. The QDs are coated with a ligand that has a negative zeta potential to prevent agglomeration while in suspension. The ligand also helps attaching the QDs to the conductive A-C electrode - this occurs between the ligand charge and the image charge that is formed in the conductive A-C electrode (screening effect). The argument for the QD array stems from the resonance behavior of their fluorescence line.

<u>The S-C cell:</u> The fact that we see capacitive amplification for QDs without white-light illumination, means that the dots were mostly thermally excited. The increase in the specific capacitance under no illumination was ca 2.5 larger than the reference (namely samples without the QDs). The increase in the specific capacitance by further illuminating the samples with white-light, cannot be attributed to the already excited QDs but to excitations of charges at the A-C electrode: firstly, the capacitance increase under white-light illumination is fairly constant irrespective of the QDs concentration, at least up to 40 μg/200 mg of A-C (Fig. 3b); secondly, the relative capacitance increase is diminishing as a function of QDs concentration (Fig. 4). This means that the illumination has less to do with the QDs (either through ionization, or energy transfer) because one may expect a larger effect while increasing QDs concentration. Thus, the added capacitance due to white-light illumination is attributed to excitations of charges (optically, or thermally, or both) in the conductive porous electrode, itself.

Charge transfer between the QDs and the A-C electrode typically appears as quenching of the fluorescence signal and in our case, should equally affect the dry and wet samples. Wet samples exhibited additional absorption and scattering compared to the dry samples and therefore, the line narrowing in wet samples is so striking. Charge transfer is affecting the life-time of the transition (estimated at sub-ns) and less its line-width [30] when the electrodes' conductively is of moderate values (namely, those values of the porous electrode). Resonance effects impact both the life-time and the line-width of the transition.

QD fluorescence dry samples: Unlike Raman spectroscopy, which details the vibrating carbon in the A-C substrates, fluorescence measurements provides details on the state of the QDs themselves and in relation with the conductive electrode. We note a substantial signal amplification for dry samples of QDs at 50 μg/200 mg of A-C, though without line narrowing (Fig. 5a). This is a linear effect that points to a characteristic distance between QDs, which is at resonance with the fluorescence line [30]. Note that signal is averaged over many such arrays and the significant one will contribute the most. If there exist an array of QDs with a relative long rang order (another words, it is large enough and occupies a relatively large area), then the fluorescence signal will also be sensitive to in-plane rotations [26]. In this case, one can maximize the signal as a function of lateral rotation and measures the line-width at the point. While most locations were not sensitive to lateral rotation, there were a few locations that do. This particular experiment was made with excitation wavelength of $\lambda_0$=532 nm of a semiconductor laser. The line-width as a function of QD titration is presented in Fig. 6. The line narrowing occurs at 50 μg/200 mg of A-C, as well. The line narrowing, measure with the $\lambda_0$=532 nm laser, is not as pronounced as in Fig. 5b (which was measured with the $\lambda_0$=488 laser on wet samples). It is also noted that the line-width is almost twice as large as the more common dry sample line-widths measured with the laser excitation of $\lambda_0$=488 nm (Fig. 5a). These resonances, therefore, allude to the formation of QD arrays. The experiment also allude to the resonating effect of the fluorescence line ($\lambda_0$=585 nm) as opposed to possible resonances of the excitation lines - both Fig. 5a and Fig. 6 point to the same QD concentration (or, to the same averaged distance between QDs) irrespective of the laser excitation line.

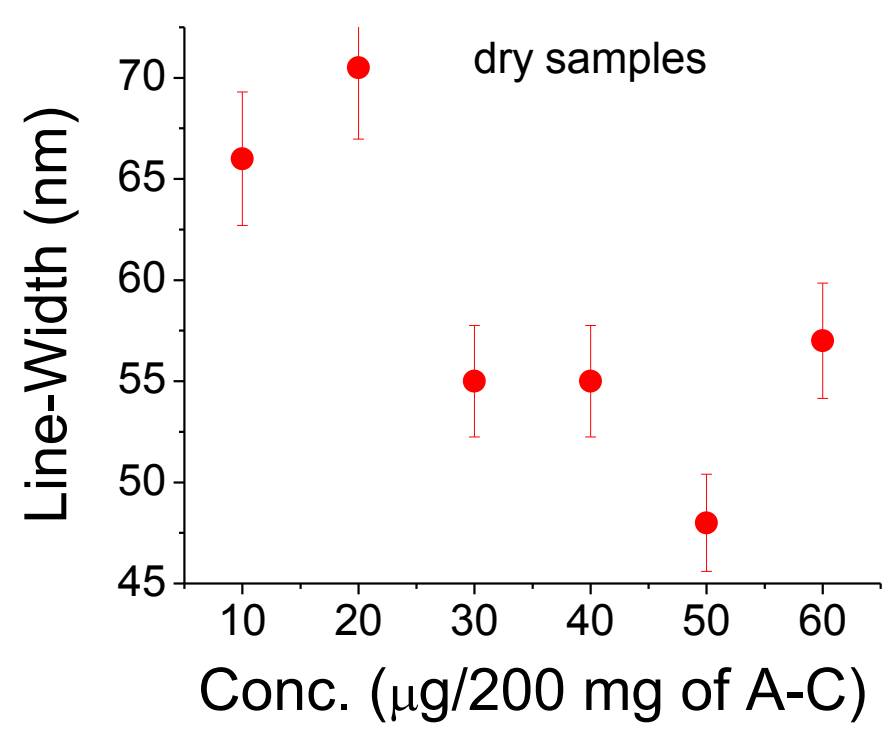

Figure 6. Line-width of fluorescence signal for dry sample. The sample was rotated in-plane to achieve maximum signal.

QD fluorescence wet samples: The wet samples (Fig. 5b) exhibited a relative signal enhancement, as well - in the 30 to 40 μg/200 mg of A-C region (not shown). In addition, it exhibited a clear line narrowing at 30 μg/200 mg of A-C. The line-width at this minimal point was smaller than the line-width exhibited by the dry samples (Fig. 5a). That line narrowing is attributed to a strong resonance effect, due, perhaps to the strong electrolyte coupling and points to fluorescence gain [30].

Theoretical considerations: For S-C, the low frequency field is applied between the electrodes, along the z-direction. In the Clausius-Mossotti relation, the macroscopic polarization is defined as, P=Np, with p the individual dipole moment (a single QD) and N the number of dipoles. This model is true when the distance between QDs is much smaller than the wavelength of interest [10-11] and is easily achievable at low frequencies of cell's operation:

$$P=N\varepsilon_0\alpha_e E_0/(1-\alpha_e C). \quad (1)$$

Here, $\varepsilon_0$ is the dielectric constant (air or electrolyte), $\alpha_e$ is the dipole polarizability, $E_0$ is the applied electric field, and C is the coupling between dipoles. Along the z-direction and between the electrodes, the coupling constant at low frequencies, $C_{LF}$, may be derived for two dimensional array of colloids, $C_{LF}\sim-2.4/\pi a^3$, with "a", the average distance between colloids. We ignore here a very small correction terms to $C_{LF}$. The polarizability may be approximated as $\alpha_e\sim 4\pi r_0^3$ if the excited dipoles are highly conductive, with $r_0$ being the QD radius (roughly 2.5 nm). Following [16] the dielectric constant is a monotonous function of the QDs concentration. The models fails at the point of transition between the organized QDs array and their amorphous phase (when the QD array is lost); this point occurs for QDs concentration of 40 μg/200 mg of A-C.

At high optical frequencies, the condition for which the propagation wavelength is much larger than the distance between QDs, is not quite met. Instead, one may use an approximation where the inter QDs separation is of the order of half a wavelength, or somewhat less. If an optical beam is normally incident onto a surface mode and is also coupled to a standing mode via a Bragg scattering in the lateral direction (the x-y direction), then resonance occurs. In the case of a quasi, 2-D QD hexagonal array of QDs that are equally separated from one another and that are attached to the electrode surface,

$$(\lambda_0/a)\cdot[(4/3)\cdot(q_1^2-q_1\cdot q_2+q_2^2]^{1/2}-n_{eff}=0. \quad (2)$$

Here, $\lambda_0$=585 nm - the fluorescence line, "a" is the inter distance between QDs and $q_1$, $q_2$ are sub-integers related to the quasi periodic structure. The effective surface guide between QDs and the electrode is $n_{eff}$. With $n_{A\text{-}C}$, $n_{air}$ equal 1.88 and 1, respectively; the effective index of the electrode/air interface is, $n_{eff}=(n_{A\text{-}C}+1)/2=(1.88+1)/2=1.44$. When introducing an electrolyte, $n_{electrolyte}\sim 1.3$ and $n_{eff}=(1.88+1.3)/2=1.59$. The electrolyte shifts the resonance condition towards the lower QDs concentration, or, put it differently, larger inter dot separation. Specifically, from 50 μg/200 mg of A-C for dry samples to 30 μg/200 mg of A-C for wet ones. The fact that the dry samples exhibited resonance at smaller inter QDs distance than the wet samples indicates that the $q_{1,2}$ are indeed sub-integers (1/2, 1/3, 1/4 etc.,). This also means that the inter QDs spacing that led to the Bragg scatterings is smaller than the propagating wavelength. Had this inter spacing larger than the propagating wavelength, the effect of the electrolyte would have been the opposite.

The inter QD distance that is attached to the electrode surface, a, may be calculated as followings: the number of clusters, N=(m)(1 g/1 L)(1/MW)(A)/(a/c). Here: m in Liters is the QDs volume in 2 mL titration vial (e.g, 30 μL for wet samples); 1 g/1 L is the stock solution of QDs (or, 1 mg/mL), MW is the molecular weight of QD including the ligand, A is Avogadro number and a/c is the number of atoms per cluster of QD; hence, $a\sim[2\ cm^3/N\ cluster]^{1/3}$. There are many possibilities here and some unknowns, e.g, MW and a/c for the QDs. Instead, one can work the formula backward, and figure out a consistent inter QD distances using Eq. 2. These come out as, a=200 nm and $q_1$=1/2 and a=170 nm and $q_1$=1/3 for wet and dry samples, respectively.

Overall, the underlined assumption of this paper was that formation of a quasi-array by functionalized QDs at the electrolyte/electrode interface led to the enhancement in both the specific capacitance in super-capacitor cells (along the z-direction), as well as the signal amplification and line narrowing (or gain) of the fluorescing line (which is polarized along the x-y plane).

## V. Conclusion

The inclusion of functionalized QDs in A-C based supercapacitors alludes to the effect of dots at close proximity to the cell's electrodes. The result is not only a large amplification of specific capacitance at some QD concentration (as measured by C-V), but also a large fluorescence enhancement with a substantial line narrowing; the latter implies significant signal gain.